\documentclass[fleqn,usenatbib]{mnras}

\usepackage{newtxtext,newtxmath}

\usepackage[T1]{fontenc}

\DeclareRobustCommand{\VAN}[3]{#2}
\let\VANthebibliography\thebibliography
\def\thebibliography{\DeclareRobustCommand{\VAN}[3]{##3}\VANthebibliography}

\usepackage{graphicx}	
\usepackage{amsmath}	
\usepackage{color, xcolor}

\newcommand{\msun}{\,${\rm M_\odot}$}
\definecolor{deeporange}{rgb}{0.91, 0.41, 0.17}
\definecolor{purp}{rgb}{0.63, 0.012, 0.98}

\title[Black hole fueling in a multiphase ISM]{Simulations of black hole fueling in isolated and merging galaxies with an explicit, multiphase ISM}

\author[A. Sivasankaran et al.]{Aneesh Sivasankaran$^{1}$\thanks{E-mail: aneeshs@ufl.edu},
Laura Blecha$^{1}$,
Paul Torrey$^{2}$,
Luke Zoltan Kelley$^{3}$,
Aklant Bhowmick$^{1}$, \newauthor
Mark Vogelsberger$^{4}$,
Rachel Losacco$^{2}$,
Rainer Weinberger$^{5}$,
Lars Hernquist$^{6}$,
Federico Marinacci$^{7}$, \newauthor
Laura V. Sales$^{8}$ and
Jia Qi$^{2}$\\
$^{1}$Department of Physics, University of Florida, Gainesville, Florida 32601, USA\\
$^{2}$Department of Astronomy, University of Florida, Gainesville, Florida 32601, USA\\
$^{3}$Center for Interdisciplinary Exploration and Research in Astrophysics, Northwestern University, Evanston, IL 60208, US\\
$^{4}$Department of Physics, Kavli Institute for Astrophysics and Space Research, Massachusetts Institute of Technology, Cambridge, MA 02139, US\\
$^{5}$Canadian Institute for Theoretical Astrophysics, 60 St. George Street, Toronto, ON M5S 3H8, Canada\\
$^{6}$Harvard-Smithsonian Center for Astrophysics, 60 Garden Street, Cambridge, MA 02138, US\\
$^{7}$Department of Physics and Astronomy "Augusto Righti" University of Bologna via Gobetti 93/2 40129 Bologna, Italy\\
$^{8}$Department of Physics and Astronomy, University of California, Riverside, 900 University Ave., Riverside, CA, 92507, USA
}

\date{Accepted XXX. Received YYY; in original form ZZZ}

\pubyear{2022}

\begin{document}
\label{firstpage}
\pagerange{\pageref{firstpage}--\pageref{lastpage}}
\maketitle

\begin{abstract}
We study gas inflows onto supermassive black holes using hydrodynamics simulations of isolated galaxies and idealized galaxy mergers with an explicit, multiphase interstellar medium (ISM). Our simulations use the recently developed ISM and stellar evolution model called Stars and MUltiphase Gas in GaLaxiEs (SMUGGLE). We implement a novel super-Lagrangian refinement scheme that increases the gas mass resolution in the immediate neighborhood of the black holes (BHs) to accurately resolve gas accretion. We do not include black hole feedback in our simulations. We find that the complex and turbulent nature of the SMUGGLE ISM leads to highly variable BH accretion. BH growth in SMUGGLE converges at gas mass resolutions $\lesssim3\times10^3{\rm M_\odot}$. We show that the low resolution simulations combined with the super-Lagrangian refinement scheme are able to produce central gas dynamics and BH accretion rates very similar to that of the uniform high resolution simulations. We further explore BH fueling by simulating galaxy mergers. The interaction between the galaxies causes an inflow of gas towards the galactic centres and results in elevated and bursty star formation. The peak gas densities near the BHs increase by orders of magnitude resulting in enhanced accretion. Our results support the idea that galaxy mergers can trigger AGN activity, although the instantaneous accretion rate depends strongly on the local ISM. We also show that the level of merger-induced enhancement of BH fueling predicted by the SMUGGLE model is much smaller compared to the predictions by simulations using an effective equation of state model of the ISM.
\end{abstract}




\section{Introduction}
 It is now well established that most massive galaxies host a supermassive black hole (SMBH) at their centres.
 Numerous observations confirm that there is a tight correlation between the mass of SMBHs and host galaxy properties such as luminosity, bulge mass and stellar velocity dispersion \citep{Kormendy1995,Magorrian1998,Ferrarese2000,Gultekin12009,mcconnell2013revisiting,kormendy2013coevolution,ReinesVolonteri2015scalingrelations,bennert2015scaling,savorgnan2016scaling}. These scaling relations indicate a connection between the SMBH activity and the evolution of the host galaxy. 
 
 SMBHs likely start as seed BHs of mass in the range $\sim 10^2-10^5$ M$_\odot$ \citep[e.g.,][]{regan2009pathways,begelman2006formation,ferrara2014initial} and grow in size primarily by gas accretion, as indicated by the \citet{Soltan1982} argument.
 During BH accretion, the infalling matter forms an accretion disk around the BH and loses angular momentum and energy via viscous dissipation. A fraction of this energy is thermalized via collisional processes resulting in the production of substantial radiation.
During high accretion this will result in very high luminosity (up to $\sim 10^{48}$ erg s$^{-1}$ ) radiation emanating from a very compact region of milliparsec size in the centre of the galaxy. These objects, called Active Galactic Nuclei (AGNs), are some of the most powerful sources of radiation in the Universe. Injection of energy and momentum feedback from AGN to the surrounding interstellar medium known as ``AGN feedback" is thought to be responsible for regulating star formation in massive galaxies \citep{benson2003shapes,DiMatteo2005,dubois2010jet,bourne2017agn-SLR}. It is also postulated to be the reason behind the observed correlations between the central SMBH mass and host galaxy properties \citep{sijacki2007unified,DiMatteo2008}. There is strong observational evidence for AGN feedback in the form of galactic outflows with velocities of the order of $\sim 1000$ km s$^{-1}$ and outflow rates (of order $\sim 1000$ M$_\odot$ yr$^{-1}$) much larger than the star formation rates \citep{Rupke2011,Sturm2011}. Such high velocities and outflow rates are difficult to achieve through stellar feedback alone \citep{Fabian2012}. Furthermore, the strongest outflow velocities are found in the central regions of the galaxies \citep{Rupke2011}. These clearly indicate the presence of AGN activity.

 In the past few decades there has been much effort to model galaxy formation and evolution with numerical simulations \citep[see][for a technical review]{VogelsbergerReview}. Numerous 
 recent cosmological hydrodynamics simulations have been able to produce observationally consistent results,  \citep[e.g.,][]{vogelsberger2014a,vogelsberger2014b,genel2014,schaye2015eagle,mcalpine2017eagle,tng-results1,tng-results2,dave2019simba} and also identify AGN feedback as the main mechanism for regulating star formation in massive galaxies \citep{teyssier2011mass,dubois2013agn,sijacki2015,TngSmbhFeedback2018}. One of the major challenges in modeling BH accretion and feedback is the large range of scales involved. Fueling SMBHs requires large inflows of gas from galactic scales to much smaller length scales such as the Bondi radius (of order $\sim$ pc). 
 Due to resolution limitations, galactic scale simulations cannot directly follow the formation of accretion disks around BHs. Instead, BH accretion and feedback are modeled using subgrid (or sub-resolution) prescriptions, which are designed to capture large scale effects of unresolved dynamical processes \citep[e.g.,][]{sijacki2015}. BHs are usually treated as sink particles that swallow nearby gas according to accretion rates calculated based on gas properties at $\gtrsim$ tens of pc - kpc 
 scales. These scales are larger than the radius of influence of BHs (distance out to which the gravitational potential of the black hole dominates the gravitational potential of the host galaxy, which is typically around a few pc to tens of pc) and hence the accretion rates are being estimated without resolving the important dynamics that ultimately delivers gas to the black hole, potentially leading to inaccuracies. 
 
 The subgrid models of the interstellar medium (ISM) can also affect the accuracy of BH accretion rates. Many hydrodynamic simulations of galaxy evolution use an effective equation of state (eEOS) approach to describe the ISM (\citealt{eEOS,GFM}; see section \ref{subsection:eEOSmodels} for a brief description). In the model we adopt, gas cooling, star formation, and stellar feedback occurring on unresolved scales are approximated via a two-phase medium in pressure equilibrium. Because these models generally do not consider gas cooling and collapse beyond a certain limit (typically $\sim 10^4$K), they produce an overly smooth gas distribution. Apart from this, the geometrical structure of the ISM is thought to play an important role in determining how the AGN feedback couples to the ISM \citep{wagner2012agncoupling,wagner2013ultrafast,bieri2017outflows,Torrey2020}. A dense and uniform ISM will allow the feedback to do more work on the environment compared to a porous ISM \citep{faucherQuataert2012physics,Torrey2020}. The geometric structure of the ISM is mostly shaped by stellar feedback and hence depends on the subgrid models of stellar evolution. Nuclear scale simulations with a multiphase ISM that resolve sub parsec scale injection of AGN feedback also show that AGNs can launch powerful gas outflows which strongly suppress nuclear star formation and BH growth \citep{hopkins2016agn}. Thus, it is important to study BH accretion and feedback using an ISM model with a well resolved multiphase structure and local injection of stellar feedback. 
 
 There exist several models in the literature that treat the ISM explicitly.
 Some of the recent ones are the Feedback In Realistic Environments (FIRE, \citealt{FIRE-2}) in the mesh-free magneto-hydrodynamics code GIZMO \citep{hopkins2015gizmo}, the \citet{Agertz2013} treatment of the ISM in the Eulerian adaptive mesh refinement code RAMSES \citep{teyssier2002ramses}, and 
 Stars and MUltiphase Gas in GaLaxiEs (SMUGGLE, \citealt{smuggle-paper}) in the moving-mesh magneto-hydrodynamics code AREPO \citep{Springel2010}. Cosmological simulations using FIRE have been able to reproduce many observed correlations of galaxy properties \citep[e.g.,][]{FIRE_mstar-mhalo,FIRE_MZR,FIRE_kennicut_schmidt,FIRE_cgm_metal_flow,FIRE_LG_dwarf_galaxies,FIRE_sfr_mstar}. Idealized galaxy simulations using the SMUGGLE model has been able to produce well resolved multiphase ISM with observationally consistent star formation rates and galactic outflows \citep{smuggle-paper}, realistic star cluster properties \citep{HuiLi2020effects,HuiLi2021formation} and Hydrogen emission brightness profiles \citep{tacchella2021h,smith2021physics} and constant density cores in dwarf galaxies \citep{jahn2021real}.
 
 Recently the FIRE model has also been used to study SMBH physics. \citet{Angles-Alcazar2017} used a gravitational torque based accretion prescription \citep{hopkins2011torque} to model BH growth in cosmological zoom simulations and showed that bursty stellar feedback limits the fueling of high redshift AGNs. \citet{anglesalcazar2021hyperLagrangian} implemented a hyper-Lagrangian refinement scheme in these simulations with which they were able to resolve the sub-parsec scale gas inflows onto the SMBH for short $\sim10\,$Myr timescales. These and related studies \citep{ccatmabacak2020black,tillman2021FIRE-QLF} of SMBHs using the FIRE model have also shown that stellar feedback and galaxy mergers play an important role in shaping the BH-galaxy scaling relations. The FIRE model was also used to simulate AGN feedback to study its impact on host galaxies and the intracluster medium \citep{Torrey2020,su2021agnjetsFIRE,wellons2022exploring}.
 
 Observational and theoretical studies have improved our understanding of black hole accretion and AGN physics. Nevertheless there are still many open questions in this field. These are some of the questions we are trying to address through this work: How does black hole fueling depend on the surrounding ISM conditions such as the availability of gas and strength of stellar feedback? What are the timescales and variability of AGN activity? To what extent do galaxy mergers trigger AGN activity?
 
In this paper we study BH fueling by simulating idealized galaxies with AREPO, using SMUGGLE to model an explicit, multiphase ISM. We investigate the nature of BH fueling in a diverse range of galactic environments and over long (few Gyr) time scales. We combine our high resolution simulations with a super-Lagrangian refinement scheme which improves the gas mass resolution in the immediate neighborhood of the BHs. This feature allows us to resolve gas dynamics near the BH on scales closer to its radius of influence. We highlight the key differences between BH accretion in the explicit ISM versus an effective equation of state ISM. After establishing the resolution convergence of our model we look at BH growth in idealized galaxy mergers. Galaxy mergers are thought to be important triggers of some AGN activity, because gravitational torques during merger can funnel gas to the central regions of the galaxies \citep[e.g.,][]{sanders1990ultraluminous, hernquist1989, barneshernquist1991, barneshernquist1996, mihoshernquist1996, springel2005modelling, hopkins2009mergersdisks, cotini2013merger, weston2016mergerAGNobs, blumenthal2018mergerInflow, ellison2019mergerAgn}. 
 The role of mergers in producing AGNs is still a debated topic, however. Some observations show that there is no strong correlation between galaxy mergers and AGN activity \citep{Grogin2005,Pierce2007,Kocevski2012,villforth2019host}. Thus, it is important to better understand mechanisms for BH fueling in both isolated and merging environments.

 This paper is organized as follows. In Section \ref{section:numerical methods} we describe the numerical methods including the subgrid models used. In Section \ref{section:results} we present the results of our simulations and analyze them. In Section \ref{section:discussions and conclusions} we discuss the implications of our results.

\section{Numerical Methods}\label{section:numerical methods}

Our simulations use AREPO, which is a moving-mesh magnetohydrodynamics code \citep{Springel2010,pakmor2016improving}.  
AREPO uses a finite volume method to solve hydrodynamics equations on an unstructured mesh defined by a Voronoi tesselation of discrete points. The mesh points are free to move with the fluid flow allowing a continuous and automatic adjustment of spatial resolution \citep[for details see][]{Springel2010}. Simulations of galaxy formation and evolution involve both dark matter and baryonic matter. In AREPO, dark matter and star particles are modelled using the collisonless Boltzmann equation coupled to Poisson's equation, while the ISM is modelled as an ideal gas following Euler's equations \citep{VogelsbergerReview}. The baryonic physics of galaxy evolution involves a large number of dynamical processes such as star formation, stellar feedback, black hole accretion, gas cooling and heating, among others, spanning a wide range of spatial and temporal scales. Because these scales are beyond the resolution limits of current simulations, these processes are implemented as subgrid models.

In this paper we will focus on simulations using the SMUGGLE model. For comparison, we will also run some of the simulations with the eEOS model. In the following sections we briefly discuss these two models.

\subsection{The effective equation of state models}\label{subsection:eEOSmodels}
In effective equation of state models of the ISM \citep{eEOS}, the small scale processes governing the dynamics of the ISM are assumed to be in an equilibrium state where the temperature of the gas can be approximated as a function of only the gas density. These models do not explicitly resolve the small scale multiphase structure of the ISM gas but instead model it as a two phase medium composed of cold gas clouds in pressure equilibrium with a hot gas. A relation of the form $T \propto \rho^\gamma$ between the local gas temperature and density is imposed (i.e., an equation of state). Stars are allowed to form stochastically from gas cells which are above (below) a given density (temperature) threshold. Due to resolution limitations, individual stars cannot be resolved and each star particle spawned from gas cells represents a stellar population. An initial mass function (IMF) is used to model stellar evolution within each star particle (for instance, the galaxy formation model of \citealt{GFM}, used in the Illustris simulations, uses the \citealt{ChabrierIMF} IMF). The energy and mass released by the stars, calculated using the IMF, is injected into a prescribed number of neighboring gas cells of the star particle in a kernel-weighted fashion. Cosmological simulations using the eEOS models use a wind model to capture galactic outflows, but that is left out of our simulations. This eEOS model has been used by Illustris \citep{vogelsberger2014a}, IllustrisTNG \citep{pillepich2018tng,weinberger2016simulating} and Auriga \citep{grand2017auriga} simulations.


\subsection{The SMUGGLE model}\label{smuggle-section}
  In order to resolve the ISM structure on scales below a few hundred parsecs, which is the scale of molecular clouds, the eEOS has to be replaced with a more detailed modelling of the cold dense gas. Current state-of-the-art galactic scale simulations have sufficient resolution to resolve gas cooling, local star formation, and feedback on much smaller scales; such simulations do not require an eEOS approach. SMUGGLE is a recently developed explicit ISM and stellar feedback model for AREPO, which is suitable for simulations with resolution of $\sim 10^4$ M$_\odot$ or better \citep{smuggle-paper}. The SMUGGLE model includes gas heating and cooling mechanisms such as low temperature atomic and molecular cooling, cosmic rays, and photoelectric heating, which can create temperatures from $\sim 10$K to $\sim 10^8$K and allow a natural development of the multiphase ISM. 
  
  SMUGGLE also models star formation and stellar feedback in a localized fashion. Star particles are allowed to form only from very cold and dense gas that is also gravitationally bound. As in \citet{smuggle-paper}, we adopt a density threshold of ${\rm 100\,cm^{-3}}$ which is in the range of densities of giant molecular clouds, compared to a threshold of ${\rm 0.13\,cm^{-3}}$ typically adopted in the eEOS model of \citet{eEOS}. Stellar feedback inputs are calculated assuming the Chabrier IMF and include supernovae (SN) energy and momentum injection, radiative feedback from young, massive stars, and energy and momentum injection from AGB and OB winds. The number of SN explosions and the mass released are calculated by sampling from a Poisson distribution in order to mimic the discrete nature of these events. SN implementation also takes into account the momentum boost due to unresolved Sedov-Taylor expansion phases.
  All stellar feedback injections happen locally into the nearest neighboring gas cells of the star particle. A cubic spline kernel is used to calculate weights and the number of neighbors is set to 32 with a tolerance of $\pm1$. 
  Galaxy properties simulated using SMUGGLE was shown to be resolution convergent at $\sim 10^4$ M$_\odot$ resolution. Details of the numerical methods and implementation can be found in \citet{smuggle-paper}.
\begin{table}
    \centering
    \begin{tabular}{|l|r|r|}
         \hline
         \hline
 Parameter & SMUGGLE & eEOS\\
         \hline
         \multicolumn{3}{|c|}{Star formation (SF) and evolution}\\
          \hline
          SF density threshold [cm$^{-3}$] & 100 & 0.13\\
          SF efficiency & 0.01 & -\\
          IMF & \citet{ChabrierIMF} & \citet{ChabrierIMF}\\
          (min, max) SNII mass [\msun] & (8, 100) & (8, 100)\\
          \hline
          \multicolumn{3}{|c|}{Black holes}\\ \hline
          Accretion factor ($\alpha$) & $10^{-5}$ & 1\\
          Radiative efficiency ($\epsilon_{\rm r}$) & 0.2 & 0.2\\
          \hline\hline
    \end{tabular}
    \caption{Key parameters in the subgrid models of star formation and evolution (gas density threshold and efficiency of star formation and IMF parameters used in stellar evolution) and BH accretion (scaling factor in Eq. \ref{eqn:bondi-rate} and radiative efficiency) adopted in our simulations using SMUGGLE and eEOS models.}
    \label{models-parameter-table}
\end{table}

\subsection{Black Hole Accretion}\label{BH-methods}
Similar to the ISM physics, it is impossible to resolve the flow of gas into the central SMBH because of the extremely small spatial scales. In our simulations the subgrid model of BH accretion is based on 
Eddington-limited Bondi-Hoyle prescription. In this model the accretion rate is given by, 
\begin{equation}
    \dot{M}_{\rm BH}=\mathrm{min}(\dot{M}_{\mathrm{Bondi}},\dot{M}_{\mathrm{Edd}}),
\end{equation}
where $\dot{M}_{\mathrm{Bondi}}$ is the Bondi-Hoyle accretion rate,
\begin{equation}\label{eqn:bondi-rate}
    \dot{M}_{\mathrm{Bondi}}=\frac{4\pi\alpha G^2 M^2_{\rm BH}\rho}{c^3_s},
\end{equation}
and $\dot{M}_{\mathrm{Edd}}$ is the Eddington accretion rate,
\begin{equation}
    \dot{M}_{\mathrm{Edd}}=\frac{4\pi G M_{\rm BH} m_p}{\epsilon_r \sigma_T c}.
\end{equation}
Here $\rho$ is the local gas density and $c_s$ is the sound speed near the BH. The factor $G$ is Newton's constant, $c$ is the speed of light in vacuum, $\epsilon_r=0.2$ is the radiative efficiency of accretion, $m_p$ is the mass of a proton, $\sigma_T$ is the Thompson cross section, and $\alpha$ is a dimensionless scaling parameter. Equation \ref{eqn:bondi-rate} is obtained by assuming spherically symmetric and steady accretion of ideal gas onto the BH. The Eddington accretion rate represents the maximum accretion rate beyond which the radiation pressure will overcome the gravitational pull on the gas. The gas density $\rho$ and sound speed $c_s$ in Eq. \ref{eqn:bondi-rate} are obtained by averaging over 64 (with a tolerance of $\pm1$) gas cells nearest to the black hole in a kernel-weighted fashion. A cubic spline kernel is used to calculate the weight of a cell as a function of its distance from the BH. The BH accretion rates do not depend strongly on the number of neighbors used in the calculation as long as it is not too large (ie, $\lesssim100$). We will often parameterize the accretion rate in terms of the Eddington ratio, $\chi_{\rm Edd} \equiv \dot M_{\rm Bondi}/\dot M_{\rm Edd}$.

The simulations in this study do not include AGN feedback. In the absence of feedback from BHs, the central gas densities can become many orders of magnitude higher than they would be in the presence of feedback. This issue is exacerbated by the SMUGGLE ISM model, which produces gas clouds with much higher densities than can be achieved in eEOS ISM models. In practice, we find that this results in 
long periods of Eddington-limited accretion
that depletes the central gas reservoir in a few hundred megayear period.

In order to study BH fueling over longer timescales without introducing a complicated dependence on the highly complex interplay between BH fueling and feedback, we adopt a non-standard approach: we use the parameter $\alpha$ to scale down the Bondi accretion rate by a factor of $10^{-5}$. This value was chosen based on the Eddington ratios of the unscaled Bondi accretion rate, which we found to have typical values of $\sim 10^5$. With this scaling of the accretion rate, we find that the Eddington ratios in our simulations vary from $10^{-12}$ to 1, with mean values (averaged over the entire simulation) of 0.05-0.3 for different simulation setups. This scaling is used only in the simulations with the SMUGGLE model, where the higher gas densities create these unphysical accretion rates. As described below, we also carry out a set of eEOS simulations for comparison. In all the eEOS simulations, $\alpha$ is set to 1, which we find yields similar average accretion rates as in the SMUGGLE model.

Historically, this $\alpha$ parameter has been used to scale {\em up} the Bondi accretion rate by an arbitrary factor \citep[often chosen to be $\sim 100$; e.g.,][]{springel2005modelling,booth2009cosmological,hayward2014galaxymergerGadget}. This was justified based on the assumption that gas densities would continue to increase on sub-resolution spatial scales, such that the Bondi rate would be underestimated in limited-resolution simulations. Essentially, this is indeed what we find in our high-resolution SMUGGLE simulations that allow high-density gas clouds to form on small scales. In reality, AGN feedback would regulate the gas density in the vicinity of the BH, such that a large reduction in the Bondi rate would not be necessary to achieve reasonable accretion rates. We note that this issue is exacerbated by the BH repositioning scheme in our runs where the BH is pinned to the local gravitational potential minimum. This issue is discussed in the next section. While the use of a scaling parameter of $\alpha=10^{-5}$ in place of AGN feedback is admittedly artificial, it enables us to achieve the primary goal of this study: conducting the first analysis of nuclear gas inflows around BHs within the SMUGGLE ISM model. The use of the $\alpha$ parameter is discussed further in Sections \ref{section:results} and \ref{section:discussions and conclusions}.\footnote{We note that the factor $\alpha$ is set equal to one in the IllustrisTNG cosmological simulations \citep{pillepich2018b}.}

\subsection{Super-Lagrangian refinement}\label{subesction-SLR}
In AREPO, gas cells are routinely refined and de-refined when their mass deviates from a fixed target mass by more than a factor of two. This ensures that all gas cells in the simulation have roughly a constant mass and hence a uniform mass resolution. In addition to this, one can also modify the refinement criterion to lower the target mass in specific regions of interest, in order to achieve a higher resolution locally with minimal increase in the overall CPU cost. Such a refinement scheme has already been used in AREPO to study cosmological gas accretion from the circumgalactic medium \citep{cgm-refinement}. We implemented an analogous scheme in AREPO to study BH accretion. We lower the target gas cell mass by a factor $F$ inside a sphere of radius $r_{\rm min}$ centred at the black hole. Beyond this radius the target mass is linearly interpolated to its original value at a radius $r_{\rm max}$ as follows, 
\begin{align}
    m(r)= \begin{cases} 
      \displaystyle\frac{m_0}{F} & r\leq r_{\rm min} \\
      \displaystyle\frac{m_0}{F}\left(1+(F-1)\frac{r-r_{\rm min}}{r_{\rm max}-r_{\rm min}}\right) & r_{\rm min}< r\leq r_{\rm max} \\
      \displaystyle m_0 & r_{\rm max}< r \label{refinement equation}
        \end{cases}
\end{align}
where $m_0$ is the uniform resolution target mass and $r$ is the distance of the gas cell from the black hole. The parameters $F$, $r_{\rm min}$ and $r_{\rm max}$ can be varied. We set $r_{\rm min}=2.86\,$kpc, $r_{\rm max}= 14.3\,$kpc and vary $F$ between 3--30 for different runs. The value of $r_{\rm min}$ has been chosen based on the largest values of the kernel radius, which is defined as the radius of a sphere centred around the BH such that the weighted number of gas cells inside the sphere is equal to the specified number of neighbours.

We note that a different refinement prescription that modifies the gas cell radius rather than the gas cell target mass was implemented in AREPO by \citet{Curtis2015}. In that scheme, the radius of the gas cells inside the refinement region is a linear function of the distance from the BH. \citet{SLR-cellradius-paper1} used this approach to formulate a modified Bondi prescription incorporating the angular momentum of the gas and showed that the angular momentum can limit BH growth and delay the quasar phase following a galaxy merger. It has also been used in several AGN feedback studies \citep{SLR-cellradius-paper2,bourne2017agn-SLR,SLR-cellradius-paper4}. The improved resolution offered by the refinement scheme was shown to make significant changes in the coupling efficiency of AGN feedback relative to identical uniform resolution simulations. However, these studies were limited by the eEOS used to model the ISM. Thus, our work using the SMUGGLE model expands on these results by explicitly modelling the formation of a multiphase ISM, as well as local injection of stellar feedback. 

Resolution of gas cells near the BH can also have an impact on the dynamics of BHs. When the mass of gas cells and other particles are comparable to the mass of the BH due to resolution limitations, two body interactions can cause the BH to wander artificially about the galactic nucleus. In order to mitigate this issue, the BH is often re-positioned to the local potential minimum at every timestep. Refining gas cells near the black hole will minimize such numerical errors and better resolve dynamical friction due to gas, making the black hole dynamics significantly more stable relative to the uniform resolution simulations. However, we find that in some of our merger simulations the position of the BHs have a few kpc scale fluctuations without the repositioning scheme even at high resolutions. Because of this, we use the repositioning scheme in all our simulations. We note that in the SMUGGLE model, especially in the absence of AGN feedback, using the repositioning scheme can lead to some enhancement of BH accretion rates due to frequent repositioning of the BH to the high density gas clouds. We find that in simulations where the BH position fluctuates by a similar amount with and without the repositioning scheme, BH masses do not differ by more than a factor of a few, indicating that our results are not dominated by the effects of the repositioning scheme. We discuss this issue in more detail in Section \ref{section:discussions and conclusions}.

\begingroup
\setlength{\tabcolsep}{4pt}
\begin{table}
    \centering
    \begin{tabular}{|l|r|r|r|r|r|}\hline \hline
        Name &${\rm M_{200}}$ & ${\rm M_{BH}}$ & ${\rm M_{bulge}}$ & ${\rm M_{disk}}$ & Disk gas \\
         & $({\rm M_\odot})$ & $({\rm M_\odot})$ & $({\rm M_\odot})$ & $({\rm M_\odot})$ & fraction\\ \hline
        MW & $1.43\times10^{12}$ & $1.14\times10^6$ & $1.50\times10^{10}$ & $4.73\times10^{10}$ & 0.16\\
        Sbc & $2.14\times10^{11}$ & $1.14\times10^6$ & $1.43\times10^{9}$ & $5.71\times10^{9}$ & 0.59\\
        SMC & $2.86\times10^{10}$ & $5.00\times10^4$ & $1.43\times10^{7}$ & $1.86\times10^{8}$ & 0.86\\ \hline\hline
    \end{tabular}
    \caption{Initial parameters for isolated galaxy simulations, as well as the progenitor galaxies for merger simulations. Columns 2 - 6 give the total galaxy mass, seed BH mass, bulge mass, disk mass and disk gas fraction of the three isolated galaxy initial conditions.}
    \label{tab:IC_params}
\end{table}
\endgroup

\begin{table*}
    \centering
    \begin{tabular}{|l|r|r|r|r|r|r|r|}
         \hline\hline
 Name & $m_\mathrm{g}$ (M$_\odot$) & $\epsilon$ (pc) & $\epsilon_\mathrm{DM}$ (pc) & $m_\mathrm{DM}$ (M$_\odot$) & $m_\mathrm{d}$ (M$_\odot$) & $m_\mathrm{b}$ (M$_\odot$) & Refinement factor(s)\\
         \hline
         resx0.1 & $2.5\times10^4$ & 45.7 & 98.6 & $2.64\times10^5$  & $2.78\times10^4$ & $2.59\times10^4$ & 3, 10, 30\\
         resx0.3 & $7.7\times10^3$ & 30.6 & 65.9 & $8.82\times10^4$ & $9.26\times10^3$ & $8.62\times10^3$ & 10\\ 
         resx1 & $2.5\times10^3$& 21.4 & 45.7 &$2.64\times10^4$  & $2.78\times10^3$ & $2.59\times10^3$ & -\\ 
         resx3 & $7.7\times10^2$& 14.3 & 30.7 &$8.82\times10^3$ & $9.26\times10^2$ & $8.62\times10^2$& -\\ \hline \hline
    \end{tabular}
    \caption{Target gas mass $m_\mathrm{g}$, softening length of baryonic particles $\epsilon$ and dark matter $\epsilon_\mathrm{DM}$ , dark matter particle mass $m_\mathrm{DM}$, stellar disk particle mass $m_\mathrm{d}$ and stellar bulge particle mass $m_\mathrm{b}$ at different resolution levels of our simulations. The last column shows the refinement factors $(F)$ used (in addition to the uniform resolution runs) at different resolution levels. In all refinement runs $r_{\rm min}$ and $r_{\rm max}$ are set to 2.86 kpc and 14.3 kpc respectively. These parameters are the same for all three initial conditions given in Table \ref{tab:IC_params}.}
    \label{resolution-table}
\end{table*}
\subsection{Initial conditions}
Our isolated galaxy simulations involve three separate initial disk galaxies: A Milky-Way type galaxy (MW), a Small Magellanic Cloud like dwarf galaxy (SMC) and a LIRG-like galaxy (Sbc). Each galaxy initial conditions (ICs) contain a dark matter halo, central SMBH, stellar bulge, stellar and gaseous disks. The dark matter halo and stellar bulge are modelled using the \citet{hernquist1990analytical} profile and the stellar and gaseous disks follow an exponential profile radially. The stellar disk has a sech$^2(z)$ vertical profile and the gaseous disk has a profile set by hydrostatic equilibrium. The stellar bulge and disk particles are tracked as distinct particle types. The IC parameters are given in Table \ref{tab:IC_params}. Our ICs are similar to the ones used by \citet{hopkins2012-IC} and \citet{hayward2014galaxymergerGadget}. We also carry out equal-mass galaxy merger simulations, using the same initial conditions as progenitor galaxies (MW-MW, Sbc-Sbc, \& SMC-SMC). In each case, the two galaxies are placed on a parabolic orbit.
The initial separation is chosen such that the pericentric passage happens at 0.57\,Gyr. The initial angular momenta of the two galaxies are at angles $(30^\circ,60^\circ)$ and $(-30^\circ,45^\circ)$ with respect to the orbital angular momentum. This orbit is chosen as a well-studied example of a strongly interacting major merger. We run simulations at four different mass resolutions (resx0.1, resx0.3, resx1, resx3). The lowest resolution level (resx0.1) has a target gas mass of $2.5\times10^4$\msun, dark matter particle mass of $2.64\times10^5$\msun, stellar bulge and disk particles with mass of $2.59\times10^4$\msun and $2.78\times10^4$\msun. In the higher resolution runs, these masses are decreased by factors of 3, 10, and 30, respectively. These parameters are summarized in Table \ref{resolution-table}. In some of the runs we combine low resolution with the refinement scheme described above to obtain a higher resolution in the central regions. For example, the resx0.1 run with a refinement factor of 30 has the same gas mass resolution in the central regions as the resx3 run. In such cases only the target gas mass in the refinement region changes, whereas the masses of DM, stellar disk and bulge particles remain unchanged.

We finally note that, owing to the smooth gas density profile in the initial conditions, an initial relaxation period is needed in our simulations before star formation and BH accretion reach steady state. We therefore do not analyze any data from the first 0.36 Gyr of each simulation. In addition, we do not allow BH accretion during the initial relaxation period.

\section{Results}\label{section:results}
\subsection{Isolated galaxies in SMUGGLE vs eEOS}
\begin{figure}
    \centering
    \includegraphics[width=\columnwidth]{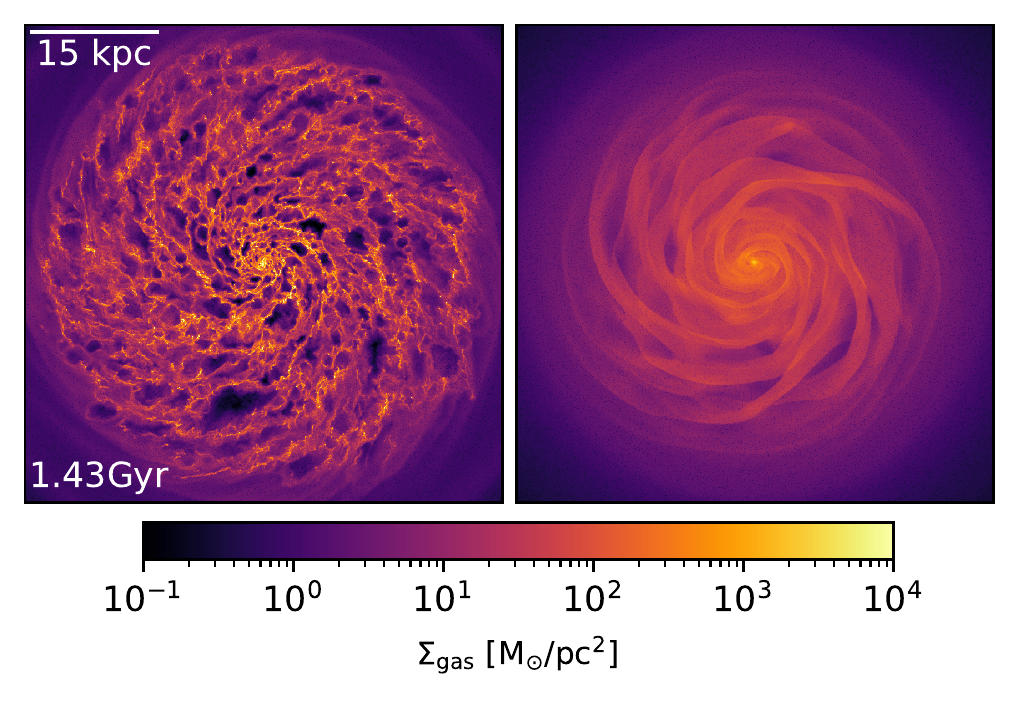}
    \caption{Face on view of gas surface density distribution in a central 5 kpc h$^{-1}$ slice of an isolated MW disk galaxy, shown at the fiducial resolution using the SMUGGLE model (left) and eEOS model (right).}
    \label{fig:smug_vs_gfm_iso2}
\end{figure}
\begin{figure}
    \centering
    \includegraphics[width=\columnwidth]{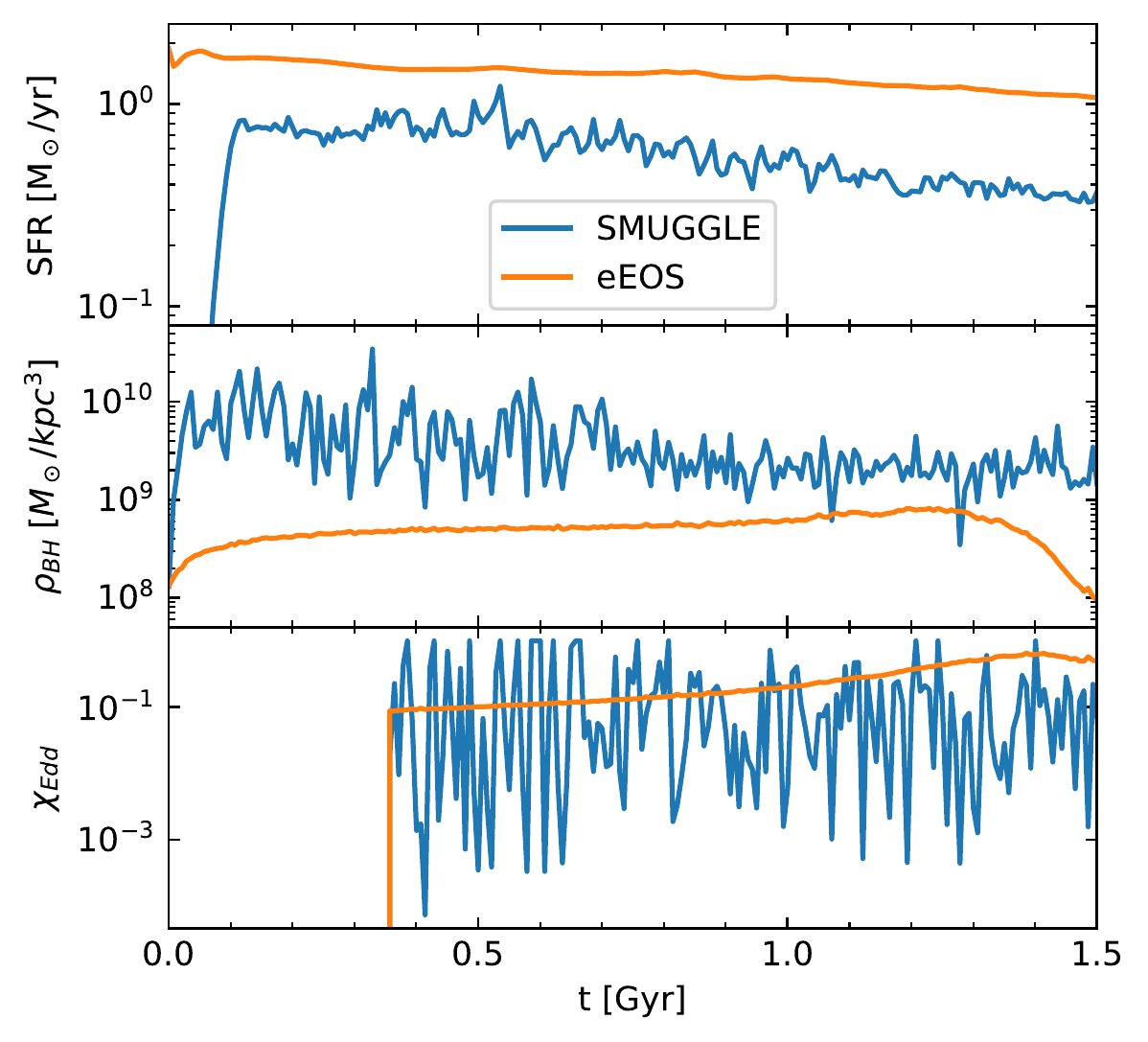}
    \caption{Total star formation rate (top), gas density near the black hole (middle) and Eddington ratio of the BH (bottom) as a function of time in the fiducial resolution (resx1) MW galaxy simulation using SMUGGLE and eEOS models. The turbulent ISM in the SMUGGLE run leads to fluctuating SFR and BH accretion rates. BH accretion is much more sensitive to the fluctuations in the ISM.
    }
    \label{fig:smug_vs_gfm_iso1}
\end{figure}
In this section we look at the main differences between isolated galaxy simulations using the SMUGGLE and eEOS model of \cite{eEOS}. Figure \ref{fig:smug_vs_gfm_iso2} shows the face-on gas density projections of our fiducial-resolution MW disk galaxy simulated using the SMUGGLE (on the left) and eEOS model (on the right). One of the main differences in the structure of the ISM is the smooth gas distributions in the eEOS ISM compared to the complex and turbulent structure of the SMUGGLE ISM. The local injection of stellar feedback in SMUGGLE leads to large ($\sim$kpc scale) low density cavities in the ISM. SMUGGLE is also able to produce gas clouds with much higher densities than the peak densities produced by the eEOS. This is due to the explicit modelling of the cold and dense phase of gas.

Figure \ref{fig:smug_vs_gfm_iso1} compares the time evolution of the total star formation rate, the gas density near the BH, and the BH Eddington ratio ($\chi_{\rm Edd} = \dot{\mathrm{M}}_\mathrm{Bondi}/\dot{\mathrm{M}}_\mathrm{Edd}$) in the two simulations. Due to the turbulent and stochastic nature of the SMUGGLE ISM, all three quantities are much more stochastically variable than in the eEOS ISM. Unsurprisingly, the environment near the BH is much more sensitive to local fluctuations in the gas density than is the global SFR. The SFR fluctuates by a factor of a few over $\sim$Myr timescales, while the BH Eddington ratio fluctuates by more than three orders of magnitude. In the eEOS ISM, the time evolution of these quantities is much smoother. The average SFR in the eEOS run is slightly higher than that of the SMUGGLE run because of the absence of stellar winds in the eEOS run. This does not make any qualitative changes to our results. The average gas densities near the BH in the SMUGGLE run are more than an order of magnitude larger than that of the BH in the eEOS run. The Eddington ratios are comparable between the two runs, however, which is a direct result of the $\alpha=10^{-5}$ scaling factor used in the SMUGGLE runs, versus $\alpha=1$ in the eEOS runs. The actual Eddington ratio of the BH (before applying the scaling factor) in the SMUGGLE run is several orders of magnitude higher than that of the eEOS run. As a result, Figure  \ref{fig:smug_vs_gfm_iso1} is useful primarily as a qualitative comparison between the two simulation setups.

\begin{figure}
    \centering
    \includegraphics[width=\columnwidth]{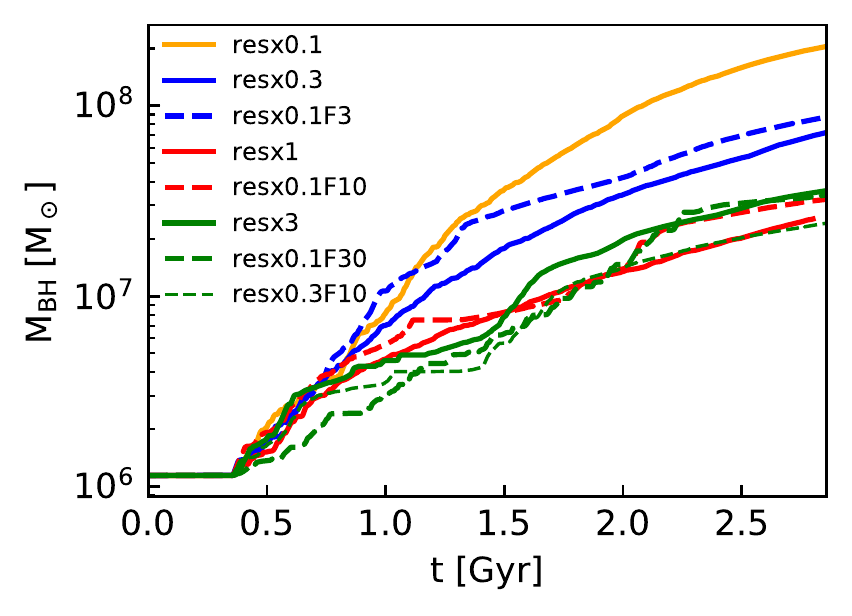}
    \caption{BH mass growth in isolated MW galaxies at four different resolution levels. Solid lines correspond to uniform resolution runs and dashed lines correspond to runs with refinement. Lines with the same color have the same resolution in the central region. For resx1 and higher-resolution runs, the BH mass growth is converged to within a factor of two. Runs with refinement are in good agreement with the corresponding high uniform resolution runs.} 
    \label{fig:MW_iso_resolution_convergence}
\end{figure}
\begin{figure*}
    \centering
    \includegraphics[width=\textwidth]{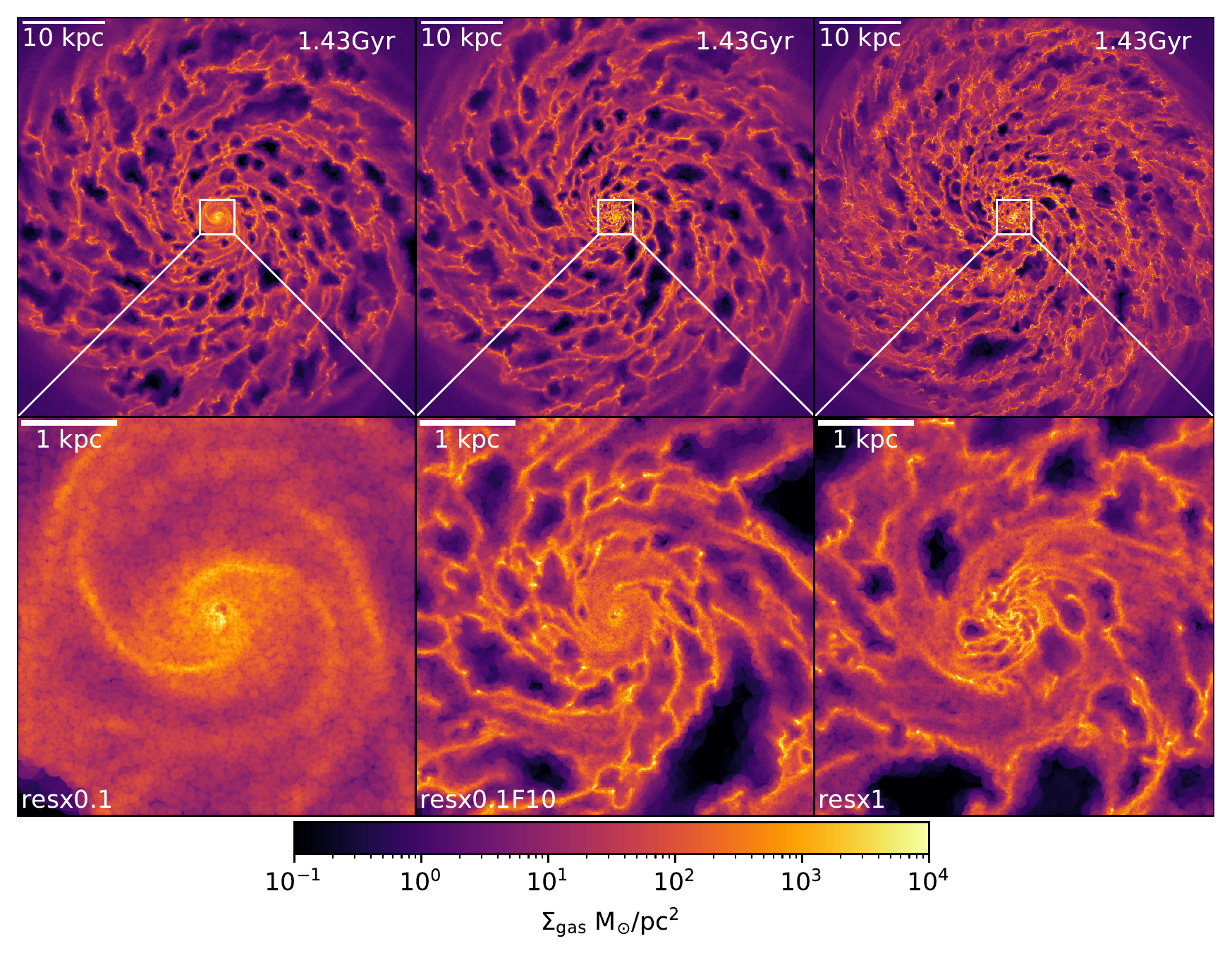}
    \caption{Top: Gas density projections in the plane of the disk, shown for a slice of 7.14 kpc thickness, for MW disk simulations at different resolutions. Specifically, each top panel shows the resx0.1 (left), resx0.1\_F10 (middle) and the resx1 (right) simulations of the MW disk at $t=1.43\,$Gyr. Bottom: Enlarged images of the central 4 kpc region of the same plots. With the refinement in the low resolution run, the central gas distribution looks very similar to that of the high resolution run.}
    \label{fig:iso_refinement_2dplots}
\end{figure*}

\subsection{Resolution convergence of BH growth in SMUGGLE}\label{section:MW_iso_resolution_convergence}
Next we look at the resolution convergence of BH mass growth in the isolated MW galaxy simulations with the SMUGGLE ISM, including the simulations with super-Lagrangian refinement around the central BH. In  Figure \ref{fig:MW_iso_resolution_convergence} we plot the mass of BHs in the MW isolated galaxies at four different resolution levels, including four uniform-resolution and four refinement runs. The dashed lines correspond to runs with refinement, and the solid lines correspond to uniform-resolution runs. In our color scheme, runs with the same gas mass resolution in the central region are given the same color. 

For the first several hundred Myr after accretion is turned on, the BHs in all simulations grow at about the same rate. After $t=1.0$ Gyr, the BH masses begin to diverge somewhat. Owing to the $M_{\rm BH}^2$ scaling of the Bondi accretion rate (Eq. \ref{eqn:bondi-rate}), we expect to see any differences in BH growth amplified over time. By comparing the uniform resolution runs, we see that the BH growth rates decrease somewhat with increasing resolution, but BH masses converge within a factor of two for the fiducial (resx1) and higher resolutions. Further increasing the resolution does not make any significant changes. The intermediate resolution run (resx0.3) has a factor of a few higher BH growth and the lowest resolution run (resx0.1) has almost an order of magnitude higher 
final BH mass compared to the fiducial resolution run. This trend is consistent with the resolution convergence of SFR (not shown here) which decreases slightly with decreasing gas mass resolution. Note that resx0.3 simulations are consistent with the lowest resolution at which SMUGGLE has previously been shown to produce reasonably well converged results for stellar and ISM evolution \citep{smuggle-paper}. 

By comparing the solid and dashed lines in Figure \ref{fig:MW_iso_resolution_convergence}, we see that the low-resolution runs with refinement produce BH growth very close to the corresponding higher, uniform-resolution run. The refinement runs are also convergent at the fiducial gas mass resolution in the central region. We have verified that increasing the refinement factor beyond 30 (we tested up to $F=100$) does not make any significant changes in the BH growth. Since at such high resolutions ($m_{\rm gas} \sim 250 \;{\rm M_\odot}$) we are reaching the limits of the stellar evolution model\footnote{An IMF is used to probabilistically sample individual stars from the star particles. If the star particles have a relatively small mass ($\sim$100\msun), this can result in artificially low number of stars at the high mass end.}, we restrict the refinement factor to a maximum of 30 (equivalent to the resx3 level). We have also included a run with a refinement factor of 10 in the resx0.3 run to show that changing both the background resolution and refinement will produce the same results, as long as the gas in the central region has the same resolution. 

In Figure \ref{fig:iso_refinement_2dplots}  we compare the face on view of the gas density distribution of the resx0.1 (left), resx0.1\_F10 (middle) and the resx1 (right) runs. In the top plots, we see that the resx0.1 and resx0.1\_F10 runs look very similar but less resolved compared to the resx1 which has finer structures. The lower plots zoom into the central 4 kpc region of the top plots. We see that the gas distribution in the central region of the lowest resolution run is very smooth on $\lesssim\,$kpc scales whereas the central region of the refinement run has more structure and looks closer to that of the higher resolution run. Thus, by using the localized, BH-based refinement scheme we can accurately reproduce the central gas dynamics of the corresponding higher uniform-resolution runs. 
\begin{figure}
    \centering
    \includegraphics[width=\columnwidth]{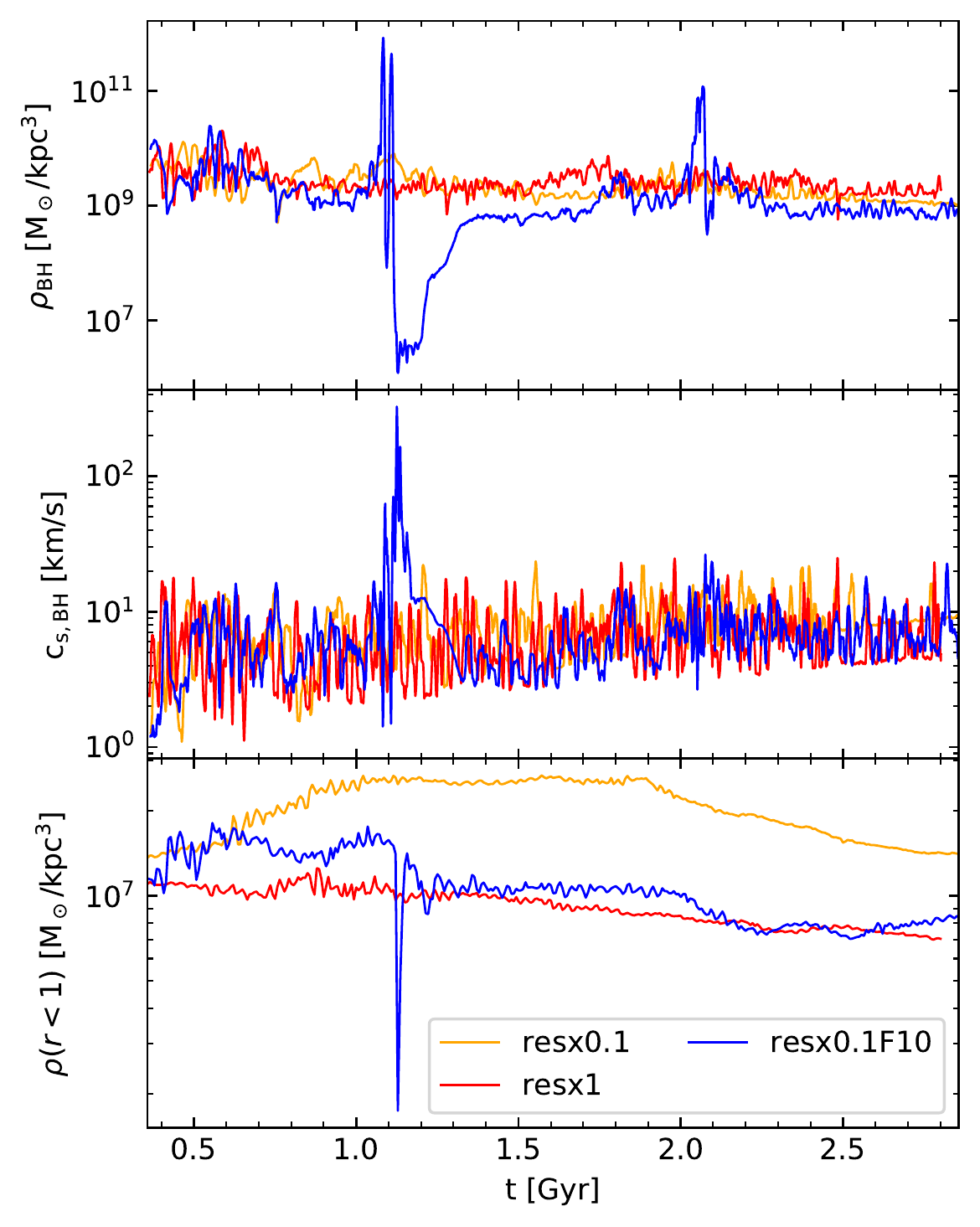}
    \caption{Gas properties near the BH in MW isolated galaxy runs at resx0.1 (uniform low), resx1 (uniform high) and resx0.1\_F10 (low refined) resolution levels. Top and middle panels show the kernel-weighted gas density and sound speed near the BH, and the bottom panel shows the average gas density within 1~kpc radius of the BH. The kernel-weighted gas density and sound speed have been averaged over 200 time-steps, which is roughly 10 Myr, and the average gas density within 1~kpc is plotted at every 7.1 Myr. The kernel-weighted gas density and sound speed of all three runs agree well. However, the average gas density in the central region of the resx0.1 run is slightly higher than the resx1 run, whereas the resx0.1F10 run is in very good agreement with resx1 run.}
    \label{fig:MW_iso_bhrho_cs}
\end{figure}

In Figure \ref{fig:MW_iso_bhrho_cs} we compare the kernel-weighted gas density and sound speed near the BH and the average gas density within a 1 kpc radius of the BH in the isolated MW simulations with resolutions of resx0.1, resx0.1\_F10 and resx1. Despite the factor of $\sim10$ difference in the final BH masses between the low resolution and the higher resolution runs, the local gas densities and sound speeds near the BH of the three runs are converged within a factor of $\sim2-3$. However, by comparing the average gas densities in the central region of the galaxies we see that the refinement run has a much better agreement with the higher resolution run compared to the low resolution run. In the resx0.1\_F10 run we see some fluctuations in the central gas density and sound speed at $\sim1.1\,$Gyr and $\sim2.1\,$Gyr. The density spike at 1.1Gyr leads to a spike in the central SFR. The SN explosions in the star particles created during the spike leads to the formation of a large cavity in the central region. This leads to a large drop in the central gas density and a corresponding increase in the sound speed which lasts for $\sim$200~Myr. We note that fluctuations of this type are not anomalous in the SMUGGLE model since the ISM is intermittently burst and turbulent \citep{torrey2017instability}. Moreover,  these quantities are calculated by averaging over a very small region of ~10-100~pc size in the central region making them very sensitive to small changes in the ISM. 

\begin{figure}
    \centering
    \includegraphics[width=\columnwidth]{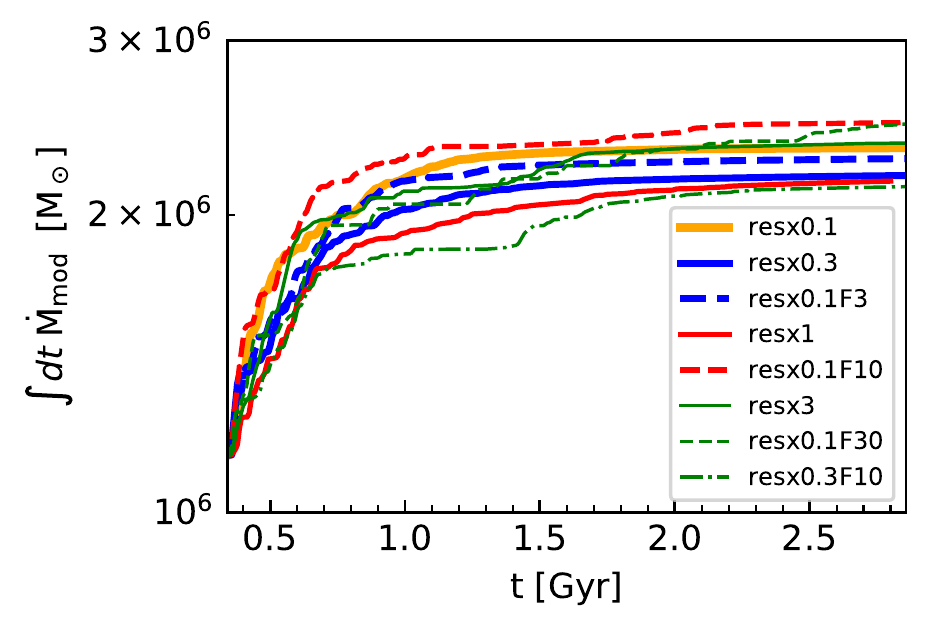}
    \caption{Gas accreted by a BH of fixed mass equal to its initial mass. Here the modified accretion rate is defined as $\dot{M}_{\rm mod}(t)=\dot{M}_{\rm BH}(t)\left(\frac{M_{\rm BH}(0)}{M_{\rm BH}(t)}\right)^2$ where $\dot{M}_{\rm BH}$ and $M_{\rm BH}$ are the Bondi accretion rates and BH mass of the runs shown in Figure \ref{fig:MW_iso_resolution_convergence}. The same line styles are used to differentiate each simulation, but the $y$-axis scale is much smaller here. All the runs have very similar mass growth, indicating that $M_{BH}^2$ dependence of Bondi accretion rate amplifies small differences over time resulting in the large differences seen in Figure \ref{fig:MW_iso_resolution_convergence}}
    \label{fig:MW_iso_ModBHMass}
\end{figure}
As noted above, the Bondi accretion rate depends on the square of the BH mass. This means that even if the average ISM properties are convergent, small differences in the instantaneous BH accretion rate can, over time, lead to fairly large differences in the BH mass. To disentangle this non-linearity and to isolate the effects of gas properties, we re-scale the accretion rate by keeping the BH mass fixed and equal to the initial mass. Thus the modified accretion rate is $\dot{M}_{\rm mod}(t)=\dot{M}_{\rm BH}(t)\left(\frac{M_{\rm BH}(0)}{M_{\rm BH}(t)}\right)^2$ where $\dot{M}_{\rm BH}$ and $M_{\rm BH}$ are the Bondi accretion rates and BH mass of the runs shown in Figure \ref{fig:MW_iso_resolution_convergence}. We then integrate this modified accretion rate with time. This gives the gas that would have been accreted by a BH of fixed mass if it were embedded in the same environment as the actual BH in the simulation. In Figure \ref{fig:MW_iso_ModBHMass} we compare this quantity at different resolution and refinement levels. Note that this is a post-processing analysis of the same simulations shown in Figure \ref{fig:MW_iso_resolution_convergence}, in a similar format. As expected we see a much better convergence here, with only a negligible difference in the total amount of gas accreted at different resolution levels (including the lowest resolution level). There is also much less gas accretion overall in this modified model--note the very different scales between Figures \ref{fig:MW_iso_resolution_convergence} and \ref{fig:MW_iso_ModBHMass}. This indicates that the differences in ISM properties determining the BH accretion rates are very small in these runs, but the non-linear dependence of the Bondi rate on the BH mass exacerbates these differences.
We would therefore expect to see even better resolution convergence in accretion models with a weaker dependence on BH mass \citep[eg,][]{hopkins2011torque,debuhr2010self,hobbs2010feeding}.
\begin{figure*}
    \centering
    \includegraphics[width=\textwidth]{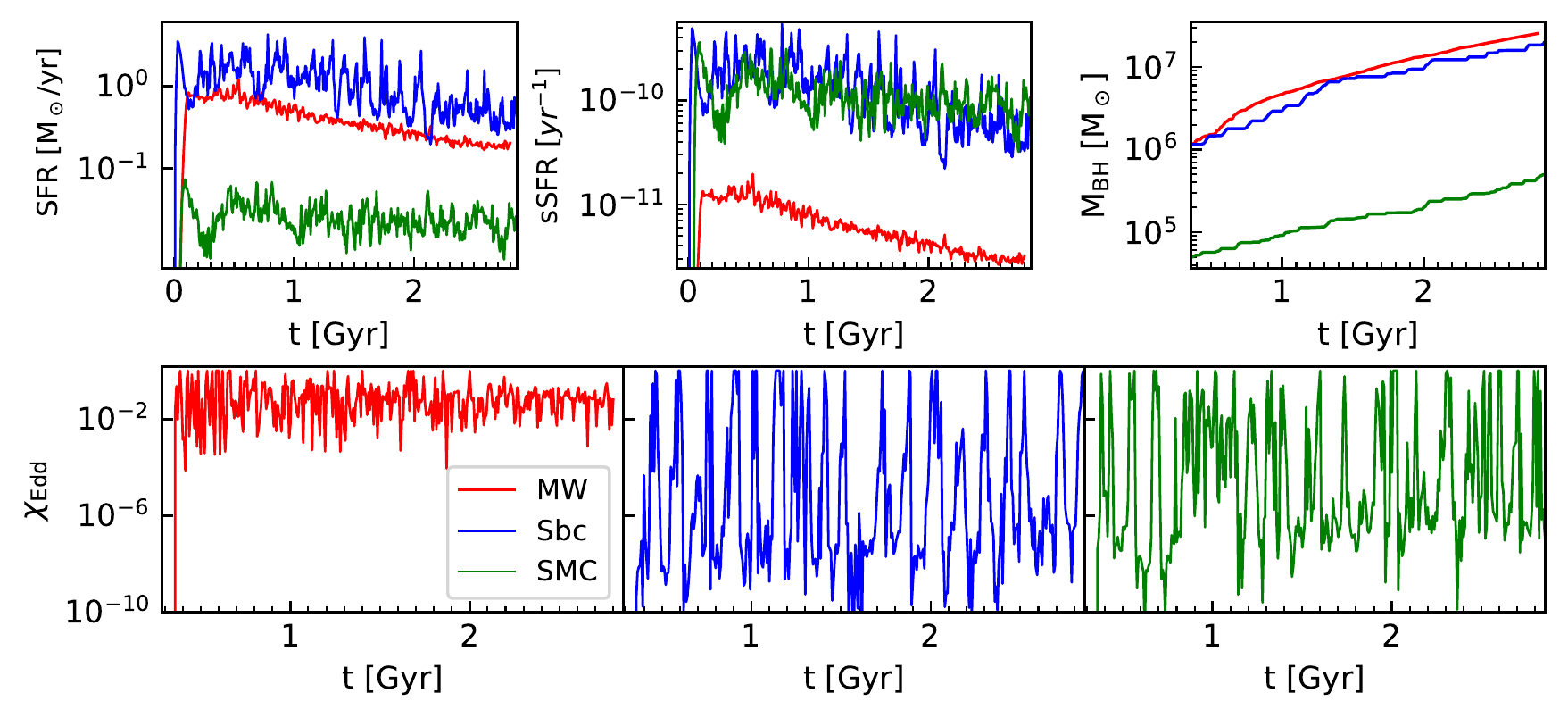}
    \caption{Top row: Total (left) and specific (middle) star formation rate and BH mass (right) as a function of time of MW (red), Sbc (blue) and SMC (green) isolated galaxies at the fiducial resolution. Bottom row: Eddington ratio of BHs in MW (left), Sbc (middle) and SMC (right) isolated galaxies at the fiducial resolution. The higher disk gas fraction of SMC and Sbc galaxies leads to more bursty star formation which results in large fluctuations in the central gas distribution. This leads to much larger fluctuations in BH accretion rates compared to that of MW.}
    \label{fig:Sbc_SMC}
\end{figure*}

The main motivation for using the super-Lagrangian refinement scheme is to attain a higher resolution near the BH without a significant increase in the run time of the simulations. The refinement runs are much faster compared to the corresponding uniform resolution runs with the differences increasing with increasing resolution. For example, the resx0.1\_F10 run is roughly 10 times faster than the fiducial resolution run (resx1) and the highest refinement run (resx0.1\_F30) is approximately 25 times faster than the highest uniform resolution run (resx3) and 5 times faster than the fiducial resolution run. Thus, in studies focusing mainly on SMBHs, using the refinement scheme can save significant computational resources while still being able to produce accurate results. We note that global properties like the SFR do not change significantly in runs with refinement.


\subsection{Sbc and SMC isolated galaxies}
We now compare results from the isolated MW disk simulation to results from the other two isolated simulations: the Sbc and SMC initial conditions. In Figure \ref{fig:Sbc_SMC} we compare the SFR, specific SFR (sSFR), BH mass and Eddington ratios of the Sbc and SMC galaxies to the MW. The top left panel shows the total SFR and the sSFR of the three galaxies. The most relevant difference between these three simulated galaxies is that the Sbc and SMC have much higher disk gas fractions compared to MW, which leads to elevated and bursty star formation with fluctuations that at times exceed an order of magnitude. As expected based on the higher gas content, the sSFRs of the Sbc and SMC simulations are very similar to each other, and both are greater than that of the MW by a factor of $\sim30$. Due to the higher disk gas fraction, the Sbc and SMC galaxies also have much higher peak gas densities near the BH compared to MW. The bursty star formation in the Sbc and SMC runs leads to large fluctuations in the central gas densities, which is reflected in the Eddington ratio evolution plots. The Eddington ratio of the BH in the MW simulation fluctuates by two orders of magnitude with an average value of $0.16$, whereas the corresponding fluctuations in Sbc and SMC span $\sim9$ orders of magnitude. These extreme fluctuations essentially mean that the BH is actively accreting for only short periods of time ($\sim$ a few Myr) when the BH resides in a high-density gaseous region, and in between these bursts of growth it is essentially quiescent. Nonetheless, the {\em average} Eddington ratios of the BHs in all three galaxies are similar (0.15 for MW, 0.13 for Sbc and 0.11 for SMC). The median Eddington ratios are 0.06 for MW, $1.8\times10^{-7}$ for Sbc and $2.2\times10^{-5}$ for SMC, owing to the much lower minimum accretion rates in those simulations. The similarity of the average Eddington ratios in Sbc and SMC indicates that the BHs spend similar amounts of time in the {\em on} state. The low values of the median Eddington ratio in Sbc and SMC are due to the large fluctuations. After BH accretion has been allowed to proceed in the simulation for a total of 2.5 Gyr,  the BH masses in MW, Sbc and SMC grow by factors of 22, 18 and 10 respectively.
\begin{figure*}
    \centering
    \includegraphics[width=\textwidth]{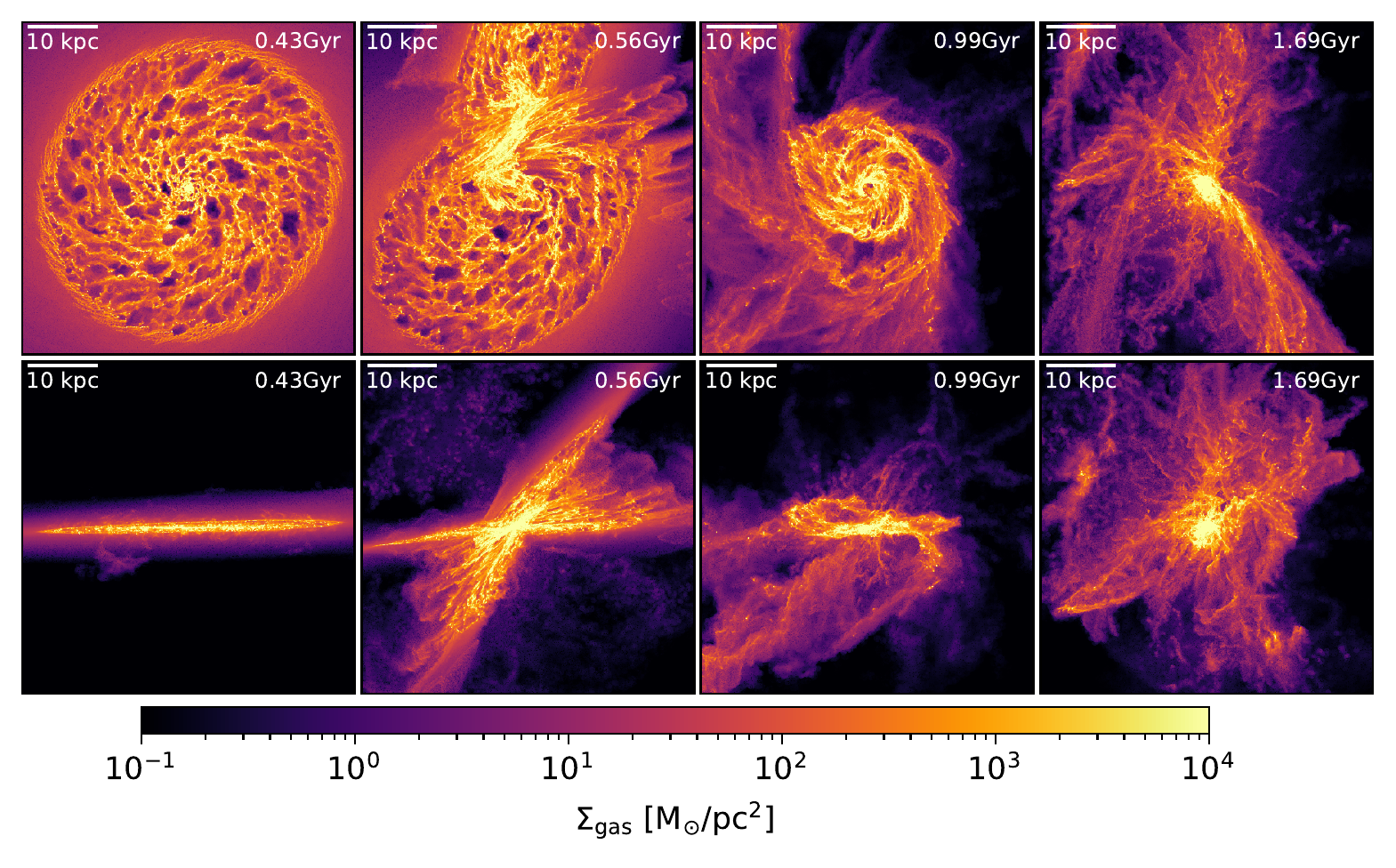}
    \caption{Gas column density projection of the resx1 MW-MW merger at 0.43Gyr (before $1^{\rm st}$ pericentre), 0.56Gyr (during $1^{\rm st}$ pericentre), 0.99Gyr (between $1^{\rm st}$ and $2^{\rm nd}$ pericentres) and 1.69Gyr (during the final coalescence). In the images shown in the top row, the line of sight is aligned with the angular momentum of one of the galaxies, and in the bottom row the line of sight is rotated by 90 degrees.}
    \label{fig:merger_2dplots}
\end{figure*}
\begin{figure*}
    \centering
    \includegraphics[width=\textwidth]{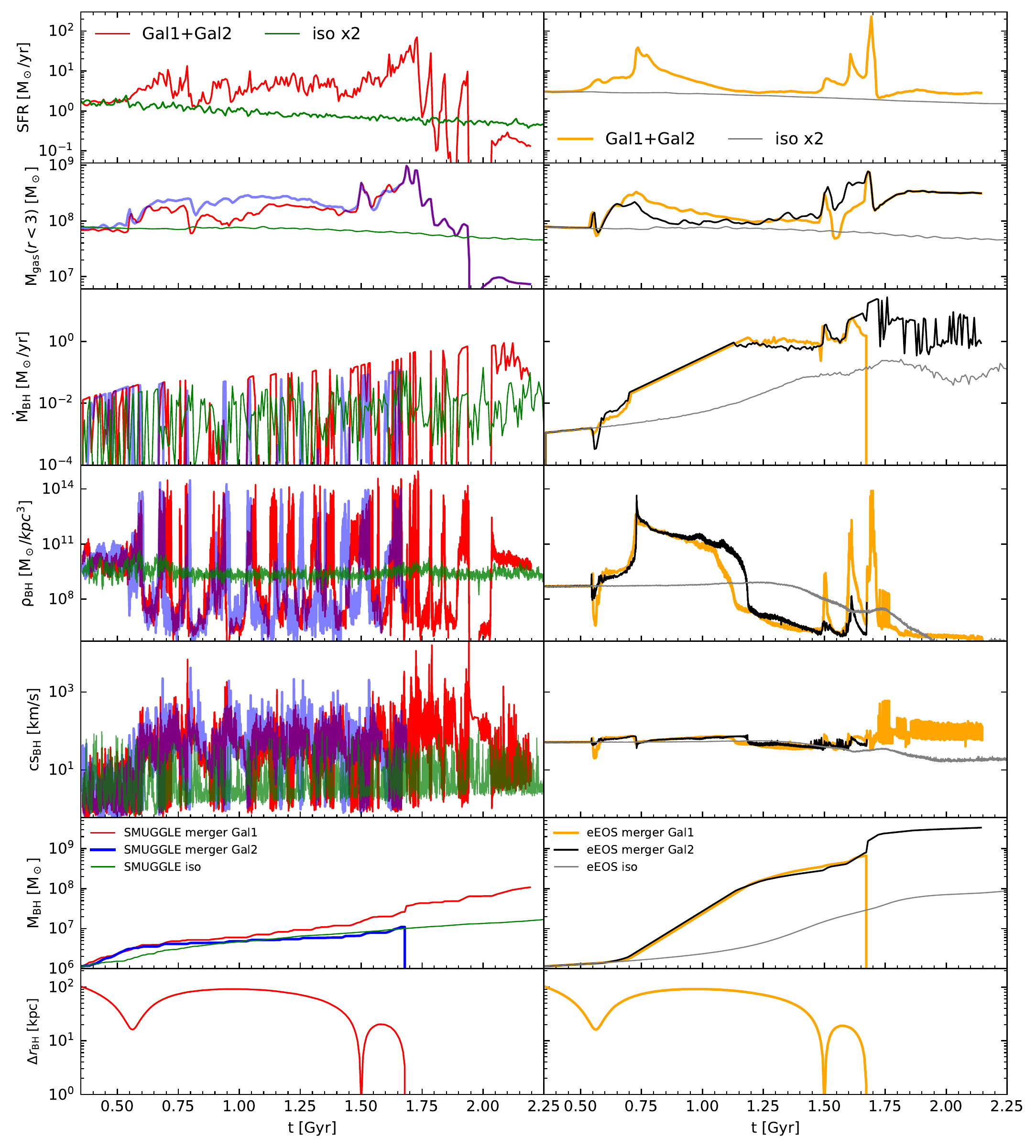}
    \caption{Top panel: Total SFR of the two MW galaxies in the merger run and twice the SFR of the isolated MW galaxy with SMUGGLE (left) and eEOS (right) models at the fiducial resolution. Panel 2: gas mass within the central 3 kpc region of the two galaxies in the merger run (red) and isolated galaxy (grey). Panel 3: Accretion rate of the BHs in the merging and isolated galaxies. Panels 4 and 5: Kernel-weighted gas densities and sound speeds near the BHs in the merging and isolated runs. Panel 6: Masses of the BHs in the merging and isolated MW galaxies. Bottom panel: Distance between the two BHs in the merging galaxies. Sound speeds and gas densities near the BHs are plotted at every timestep of the simulation. All other quantities are plotted at 7.1 Myr intervals. The interaction of the galaxies trigger gas inflow towards the central region resulting in elevated SFR and BH accretion rates. Due to the large fluctuations in the central gas distribution in the SMUGGLE run, the merger induced enhancement of BH mass growth is much smaller than that of the eEOS runs.}
    \label{fig:merger_All_props}
\end{figure*}

\subsection{Idealized Mergers}

\begin{figure*}
    \centering
    \includegraphics[width=\textwidth]{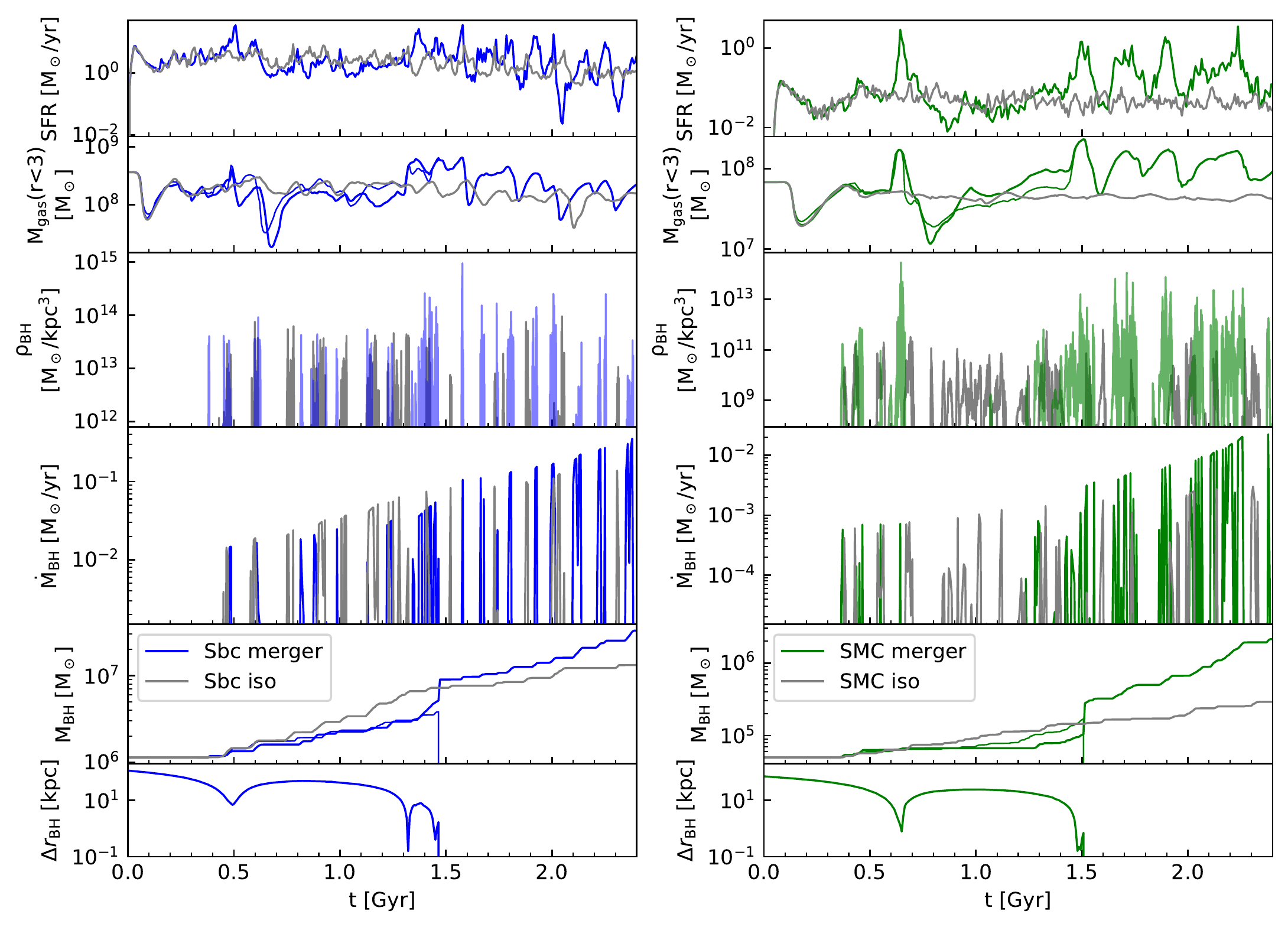}
    \caption{Top panel: Total SFR of the merging galaxies and twice the SFR of the isolated galaxy with Sbc on the left and SMC on the right. Panel 2: Gas mass within 3kpc of the two merging galaxies and the isolated galaxy. Panel 3: Kernel-weighted gas density near the merging and isolated BHs at each time-step of the simulations. Panels 4 and 5: Accretion rates and masses of the BHs in the merging and isolated galaxies. Panel 6: Distance between the two BHs in the merging galaxies. Gas densities near the BHs are plotted at every timestep of the simulation. All other quantities are plotted at 7.1 Myr intervals. In both SMC and Sbc mergers, strong stellar feedback after the $1^{\rm st}$ pericentric passage expels gas from the central region resulting in lower BH accretion until the $2^{\rm nd}$ passage. The final coalescence leads to gas inflows resulting in elevated accretion rates. The overall enhancement of BH growth in SMC is similar to that of MW, whereas Sbc has only a very small enhancement.}
    \label{fig:merger_sbc_smc}
\end{figure*}

\subsubsection{MW-MW mergers}
In this section we show the results of our idealized merger simulations, focusing first on mergers between the MW progenitor galaxies. As with the isolated runs, we have verified that the super-Lagrangian refinement merger simulations are in reasonably good agreement with the corresponding uniform high resolution runs. Accordingly, we do not present further detailed comparison between the refinement and uniform-resolution merger simulations. Figure \ref{fig:merger_2dplots} shows both face-on and edge-on gas column density projection of the fiducial-resolution SMUGGLE MW merger at four different epochs: before the first pericentric passage, during the first pericentric passage, between the first and second pericentric passages and during the final coalescence. In the top images the line of sight is aligned with the angular momentum of the galaxy being plotted (and the BH is at the centre of the box). In the bottom plots the line of sight is rotated by 90 degrees. By comparing the first and third snapshots, we see that the close interaction between the galaxies strips some of the gas from the outer regions of the galaxies. Nonetheless, 
the galaxies still retain the disk structure and higher gas densities in the central region, which is clear in both the edge-on and face-on views. The fourth snapshot shows that the final coalescence completely deforms the shape of the galaxies. A large fraction of the gas ($\sim35\%$) in the galaxies is now clustered in a very small ($\sim5\,$kpc) region at the centre with densities decreasing very rapidly with distance from the centre. After this snapshot strong feedback from SN explosions expel most of the gas from the central region of the galaxies. 

In Figure \ref{fig:merger_All_props} we compare the MW-MW merger simulations at the fiducial resolution (resx1) with the SMUGGLE (on the left) and eEOS (on the right) models. In the bottom panel we plot the distance between the BHs in the two galaxies. In both models, the galaxies go through two pericentric passages before the final coalescence\footnote{The BHs of the two galaxies are merged when they are within the neighbor search radius.} at $t\sim1.7\,$Gyr. In the first and second panels we show the total SFR and the gas content in the central 3 kpc of the galaxies in these runs. The green and grey lines correspond to the SMUGGLE and eEOS isolated galaxies at the fiducial resolution.  
The SFRs of the isolated galaxies are multiplied by two for easier comparison. 

In the SMUGGLE run, after the pericentric passages and the coalescence, there is a spike in the central gas content of the merging galaxies relative to the isolated galaxies. This indicates that the interaction between the merging galaxies drives an inflow of gas towards the nuclear region. The increase in nuclear gas content is also correlated with a spike in the star formation rates of these galaxies due to increased gas densities. After the first pericentric passage the SFR of the merging galaxies goes up by a factor of $\sim10$ followed by another increase by a factor of $\sim10$ after the coalescence. After $\sim1.7\,$Gyr, the SFR fluctuates by 2-3 orders of magnitude and decreases rapidly as most of the central gas is expelled by stellar feedback. In the eEOS merger we see a similar increase in the nuclear gas content and SFR after the pericentric passages and coalescence, but the SFR varies in a smooth fashion, whereas the SMUGGLE galaxies are significantly more bursty. After the final coalescence in the eEOS runs there is no secular decrease in the SFR and central gas content.

In rows 3-6 of Figure \ref{fig:merger_All_props} we compare the time evolution of the BH accretion rates, gas density and sound speed near the BHs, and the BH masses in these runs. We focus first on the SMUGGLE results. The gas densities near the BHs in the SMUGGLE merger run stay roughly the same with small fluctuations until 
the first pericentric passage. After the passage, the peak densities increase by 2-3 orders of magnitude, followed by a further increase by a factor of few after the final coalescence. However, the strong stellar feedback due to the elevated and bursty star formation after the pericentric passage leads to the formation of large low density cavities in the central regions of the galaxies. This results in large fluctuations (4-5 orders of magnitude) in the central gas density and limits the average gas density as the BHs end up spending significant amounts of time in these cavities. The maximum central sound speeds also increase by 2-3 orders of magnitude, as the same feedback that drove the formation of the cavities also heated the gas. Note that the minima of the sound speed are correlated with the peaks of density. Additionally, the minimum central sound speeds in the merger run are lower by a factor of a few relative to the equivalent isolated galaxy. 

The higher densities and lower sound speeds lead to large spikes in the Bondi accretion rates, which are suppressed by the $\alpha$ scaling factor and by the Eddington limit. During the density peaks, the BHs accrete near the Eddington limit for short durations, in between which the accretion rates are low. Between the first pericentric passage and the final coalescence there is a factor of few increase in the average accretion rates. During and after the coalescence, the accretion rates increase further, to a factor of $\sim 10$ above their initial average values. Thus, although local stellar feedback modulates the BH accretion rate on short timescales as in the isolated galaxy, here the interaction between MW-type galaxies drives cold gas to the galactic nuclei, thereby enhancing the total BH growth. In the fiducial-resolution SMUGGLE runs the total BH mass growth over a period of $\sim1.9\,$Gyr in the merger run is enhanced by a factor of $\sim4$ relative to that of the isolated run.

In the eEOS merger, there is a steady increase in the gas density near the BH after the first pericentric passage. 
As in the SMUGGLE merger, this results from the nuclear inflow of gas driven by gravitational perturbations. But in this case, the smooth gas distribution in the central region of the galaxies (as seen in Figure \ref{fig:smug_vs_gfm_iso2}) yields steady, Eddington-limited accretion for $\sim500\,$Myr, after which the gas densities and accretion rates decrease as the galaxies run out of gas in the central region. During the second pericentric passage and the final coalescence, more gas in funnelled into the central region, and as a result the central gas density and accretion rates spike again. The central sound speed increases slightly after the pericentric passages and the coalescence. Although the time averaged central densities are comparable in both the eEOS and SMUGGLE mergers, the factor of $\sim10$ difference in the minimum sound speeds make a significant difference in the accretion rates in the two models due to the $c_s^{-3}$ dependence of the Bondi accretion rate. The Eddington ratio of the BHs in the eEOS merger has an average value of 0.65 and a median of 0.59 during 250 Myr period immediately after the first passage. During the coalescence the average is 0.63 and median is 0.8. In the SMUGGLE merger the average Eddington ratio is 0.23 during both phases, but the median is lower during these periods ($\sim 2-3 \times 10^{-6}$), again reflecting the large fluctuations between active accretion episodes and inactive periods. 
After $\sim2\,$Gyr the total BH growth in the eEOS merger run is enhanced by a factor of $\sim40$ relative to the eEOS isolated run, which is an order of magnitude larger enhancement than that in the SMUGGLE run. 

Note that $\alpha=1$ in the eEOS runs, meaning that no scaling factor is applied to the Bondi rate. As indicated by the density and sound speed plots (Figure \ref{fig:merger_All_props}), the BH growth in the SMUGGLE runs would be much larger than that in the eEOS runs if no scale factor were applied to the Bondi rate. We also note that since most of the BH growth in the eEOS merger run happens during the steady, Eddington-limited accretion, the final BH mass is not expected to change significantly for $\alpha$ in the range $\sim 0.001-1$. It is difficult to predict exactly how these comparisons would change in SMUGGLE runs with $\alpha=1$ that also included some form of BH feedback. Depending on the nature and strength of the feedback, and the galactic environment, we might see either a larger or a smaller difference in BH growth in merging galaxies relative to the isolated galaxies. 

Our results from the eEOS run are in general consistent with past studies of BH growth during mergers \citep[e.g.,][Thomas et al. in prep]{blecha2011recoilingBH,debuhr2011BHgrowth-merger-eEOSm}. In the SMUGGLE model the large scale (few kpc) gas inflow driven by the interaction of the galaxies is very similar to that in the eEOS model. But the fueling of BHs that happens at much smaller scales ($\sim10\,$pc) is dramatically different due to the stochastic stellar feedback and the presence of cold and dense gas clouds. Even though the BH fueling in the SMUGGLE run is enhanced during the merger, the level of enhancement is much more modest compared to that in the eEOS model.

\subsubsection{Sbc and SMC merger simulations}

In Figure \ref{fig:merger_sbc_smc} we show the SFR, BH accretion rate, BH mass, and separation between the BHs in the Sbc-Sbc (left) and SMC-SMC (right) merger simulations with the SMUGGLE model. In the SMC run the SFR and nuclear gas content spikes for a short duration at the time of the first pericentric passage and decreases afterwards. This is different from the case of MW merger, where the SFR and central gas content increases almost steadily following the first passage. As seen in Figure \ref{fig:Sbc_SMC}, star formation is significantly more bursty in the SMC run compared to the MW. The close interaction of the galaxies during the first passage leads to a further increase of SFR and hence stronger stellar feedback. Strong winds driven by the feedback expel a large amount of gas from the central region of the galaxies, thereby limiting the gas inflow and lowering the SFR. This is seen as the sudden drop of ${\rm M_{gas}}$ at $\sim0.8\,$Gyr. This results in lowered accretion rates until the second pericentric passage. The final coalescence brings more gas to the central region of the galaxies. After the coalescence, the central gas content increases by a factor of $\sim5$ and the global SFR increases by roughly two orders of magnitude with large fluctuations over $\sim100\,$Myr timescales. This increases the local gas density near the BHs and consequently the time averaged accretion rate increases by an order of magnitude. The final BH mass in the SMC merger at $t=2.4$ Gyr is $\sim4.3$ times larger than twice the final BH mass in the isolated galaxy. The enhancements in the SFR and the BH accretion rates after the final coalescence in the SMC merger are similar to that of the MW merger.

In the Sbc merger the global SFR and nuclear gas content behave very similarly to SMC until the second pericentric passage due to the similar bursty star formation of Sbc. The spike in SFR during the first passage is smaller compared to that of SMC. The expulsion of gas by stellar feedback from the galactic centres after the first passage slightly lowers the accretion rates. After the second pericentric passage and the final coalescence the SFR and nuclear gas content increase by less that an order of magnitude and a factor of $\sim3$ respectively. The time averaged BH accretion rates increases only by a factor of few after the coalescence, and the total BH mass growth is enhanced only by a factor of 1.25 relative to the isolated run. These enhancements are much smaller than the corresponding enhancements of SFR and BH accretion rates  in the MW and SMC merger.

Thus, the interaction of the galaxies enhance the BH accretion rates in all three merger simulations. The effect of the first pericentric passage on the SFR and BH accretion rates is more pronounced in the MW merger, whereas the SMC and Sbc mergers have virtually no nuclear gas inflows at this stage. After the coalescence all three systems become significantly more bursty, with the MW merger showing the largest fluctuations in SFR. BH accretion rates also increase, in correlation with the SFR. The accretion rates in Sbc and SMC mergers are lower that that of the isolated runs until the coalescence and increases to a higher value after the coalescence resulting in an overall enhancement of BH mass growth. Thus, our results indicate that galaxy mergers can play an important role in triggering AGN activity, in agreement with previous work, but we also find that the amount and timescale of merger-induced BH fueling episodes depend strongly on the nature of the surrounding ISM.

\section{Discussion and Conclusions}\label{section:discussions and conclusions}
In this paper we have investigated the nature of gas inflows and BH fueling in AREPO hydrodynamics simulations using the multiphase ISM and explicit stellar evolution model SMUGGLE. Our simulation suite included initial conditions for three different progenitor galaxies: a gas-poor MW type galaxy and two gas rich galaxies (Sbc and SMC). These were each evolved in isolation, and also in a set of idealized galaxy merger simulations, over long timescales of $\sim$ 2-3 Gyr. Most of our simulations 
resolve gas dynamics at $\sim10-100\,$pc scales.

We find that the clumpy and turbulent ISM of the SMUGGLE model results in stochastic fueling of BHs with orders of magnitude fluctuations in the accretion rates over $\sim\,$Myr timescales. Star formation is also stochastic in the SMUGGLE model, but we find that BH accretion rates are much more sensitive to the stochastic variations in the local ISM density. Gas-rich galaxies with bursty star formation have larger fluctuations in the BH accretion rates due to the formation of large low density cavities in the ISM. The stochastic nature of BH accretion shown in our work is similar to the behavior seen in FIRE simulations \citep{Angles-Alcazar2017} with a gravitational torque based accretion prescription \citep{angles2017torquemodel}, although the type of galaxies in their study were different from ours. They also found bursty BH accretion modulated by stellar feedback, especially at high redshifts.

By varying the resolution levels, we have demonstrated that the Bondi-Hoyle accretion model coupled to SMUGGLE is resolution convergent for gas cell masses $\lesssim3\times10^3\mathrm{M_\odot}$. The BH masses after 2.5 Gyr agree to within a factor of $\sim4$ for all except the lowest resolution run. When the $M_{\rm BH}^2$ dependence is scaled out of the Bondi accretion rate to compare the amount of gas accreted by BH of fixed mass, all the runs agree within a factor of $\sim1.5$. Additionally, we implemented a super-Lagrangian refinement scheme which increases the gas mass resolution in the immediate neighborhood (within 2.86 kpc) of the BHs. Using this scheme in the lowest resolution simulations ($m_{\rm gas} = 2.5\times10^4\mathrm{M_\odot}$) produced central gas dynamics and BH growth in good agreement with that of the high-resolution runs. This allows high resolution studies of BH dynamics with only a minimal increase in CPU cost. Isolated galaxy runs with refinement factors of 10 and 30 were approximately 10 and 25 times faster than the corresponding high uniform resolution runs. Often, the highest resolution achieved by cosmological and zoom in simulations is comparable to the lowest resolution considered here.   
This implies that using the refinement scheme in these simulations can accurately resolve BH accretion where the BH growth will otherwise be overestimated due to the low resolution. However, further testing is required to ensure the validity of the refinement scheme for refinement factors above 30. 

In our isolated disk simulations, we find that the higher gas fraction of Sbc and SMC galaxies leads to much more stochastic star formation and BH fueling relative to that of the MW galaxy. The BH in the MW-like simulation has an average Eddington ratio of 0.16 with roughly two orders of magnitude fluctuations over Myr timescales whereas the BHs in Sbc and SMC like galaxies have average Eddington ratios of $\sim$ 0.1 with 9 orders of magnitude fluctuations. Thus, in the latter cases the BH is essentially switching back and forth between an active fueling state and a quiescent state, dictated by whether it is instantaneously located in a high-density gas cloud or a low-density region evacuated by local stellar feedback. In reality, the accretion rate should vary more gradually as the accretion disk is consumed by the BH over the viscous timescale, and AGN feedback would further modulate the fueling environment. Nonetheless, these results demonstrate the dramatic impact that the multiphase ISM can have over short timescales on the cold gas supply available to the central BH.

We also studied BH fueling during galaxy mergers. The impact of different stages of the merger on BH accretion depends strongly on the type of galaxies. In the MW merger, the first pericentric passage leads to large inflow of gas to the central region of the galaxies. This increases the peak gas densities near the BHs by several orders of magnitude while also leading to large fluctuations due to the bursty stellar feedback. The net effect is an increase in average accretion rates. In the Sbc and SMC mergers, owing to their high sSFR, strong stellar feedback following the first pericentric passage expels a large amount of gas from the central region of the galaxies resulting in lower accretion rates. Thus the first pericentric passage produces opposite effects on BH fueling in Sbc and SMC runs compared to the MW run. However, the second pericentric passage and the final coalescence leads to inflows in all three systems resulting in enhanced BH accretion. The inflows in MW and SMC mergers are much stronger compared to the inflows in the Sbc merger. The total BH mass growth in the MW and SMC merger runs is enhanced by a factor of $\sim4$ relative to that of the corresponding isolated runs whereas in the Sbc merger run there is only a factor of $\sim1.25$ enhancement.
Thus the overall enhancement of BH growth depends strongly on the initial conditions. We also compared our SMUGGLE runs with simulations of the same initial conditions using the eEOS model, the ISM and stellar evolution model used in Illustris, IllustrisTNG and Auriga simulations. We find that the merger induced enhancement of BH fueling in the SMUGGLE model is much smaller compared to that in the eEOS model. This is because of the bursty stellar feedback in the SMUGGLE runs which results in high variability at short timescales of the central gas distribution counteracting the inflows. In the MW merger with the eEOS model, the smooth ISM and the inflow of gas  leads to  a steady increase in BH accretion rate with a factor of $\sim40$ enhancement in the total BH mass growth relative to the isolated runs.

These results imply that major mergers of the type of galaxies considered here can trigger AGN activity. However, simulations that do not explicitly model stellar feedback and resolve the multiphase structure of the ISM can significantly overestimate merger induced enhancements in AGN activity.

Although the initial conditions of our simulations represent diverse types of galactic environments, to reach more general conclusions regarding the nature of BH accretion flows in a multiphase ISM, more exhaustive study of the parameter space would be needed. Such type of studies will be easier to perform using simulations of a cosmological volume, an avenue that we aim to pursue in future work.

The subgrid prescriptions for BH accretion, feedback, and dynamics are also important areas for future work. The simulations presented here do not include any AGN feedback, in order to focus on nuclear gas inflows separately from their highly nontrivial coupling to AGN outflows. Because the explicit ISM model allows gas to condense to very high peak densities, particularly in galactic nuclei, the Bondi accretion model produces unrealistic BH mass growth over time in the absence of AGN feedback. As discussed in Section \ref{BH-methods}, in order to simulate long periods (several Gyr) of galactic evolution in isolated and merging systems,  we used the $\alpha$ parameter in the Bondi accretion formula to artificially scale down the accretion rates by a large factor ($10^{-5}$). With this scaling, the average and median values of the uncapped Eddington ratios in the isolated disks are  $(0.31, 0.06)$ in MW, 
$(0.75, 2.2\times10^{-5})$ in SMC and $(110, 1.8\times10^{-7})$ 
in Sbc runs. In the merger runs the average and median Eddington ratios are 
$(260, 2.5\times10^{-6})$ in MW, $(23, 1.1\times10^{-5})$ for SMC and 
$(150, 1.7\times10^{-7})$ in Sbc. Given the stochastic, highly variable nature of BH accretion within the SMUGGLE ISM model, these appear to be reasonable values for the range of AGN simulated here \footnote{We emphasize that the accretion rates are in fact capped at the Eddington limit in our simulations; the ``raw" Bondi-Hoyle rates are presented here for reference.}.
In reality, BH growth is expected to be self-regulated through AGN feedback. In future simulations, we plan to study this self-regulated growth directly. 

BH dynamics, including dynamical friction, cannot be fully resolved even in the highest-resolution regimes of our simulation suite. To prevent spurious wandering of the BH due to numerical noise, we therefore employ the common approach of repositioning the BH on the potential minimum at every timestep. As mentioned in section \ref{subesction-SLR}, we find that this BH repositioning scheme can have some significant effects on BH dynamics in the SMUGGLE ISM. Because this model directly traces the formation of high-density gas clouds and local stellar feedback injection within a clumpy, multiphase ISM, the gas distribution varies dramatically over short time and spatial scales. As a result, 
the repositioning scheme has a tendency to localize the BH within high density gas clouds, which can lead to some artificial enhancement of the accretion rate. However, we find that not using the repositioning scheme in merger simulations can artificially suppress the BH accretion rates, particularly in the post-merger stage, as the BH is subject to spurious dynamical kicks from comparable-mass particles. 
In future work, it will be useful to consider a more realistic dynamical friction prescription in place of this 
repositioning scheme. However, we note that this issue will also be mitigated by the inclusion of AGN feedback, as the feedback will quickly disperse the high density gas clouds and decrease the gas density in the central region of the galaxies.

The interplay between BH fueling, star formation, and stellar feedback is a crucial but complex aspect of BH and galaxy co-evolution. We have undertaken a novel numerical study of BH fueling in an explicit, multiphase ISM. We implemented a novel super-Lagrangian refinement scheme and studied gas inflows onto BHs over long timescales and in diverse merging and isolated galactic environments. Our results clearly demonstrate the profound impact that the local ISM can have on BH fueling rates and duty cycles, while confirming previous findings that global perturbations (such as galaxy interactions) also modulate nuclear gas inflows and trigger AGN activity. Our work lays the groundwork for future studies of BH/galaxy co-evolution, including studies that also consider the role of AGN feedback within an explicit, multiphase ISM. 

\section*{Acknowledgements}
LB acknowledges support from Cottrell Scholar Award \#27553 from the Research Corporation for Science Advancement. LB and PT acknowledge support from the NASA Astrophysics Theory Program award \#80NSSC20K0502.  
PT acknowledges support from NSF-AST 2008490. 
LVS is grateful for support from the NSF- AST 1817233, CAREER-1945310 and NASA ATP-80NSSC20K0566 grants. RW is supported by the Natural Sciences and Engineering Research Council of Canada (NSERC), funding reference \#CITA 490888-16. The simulations in this work were performed on the supercomputer HiPerGator at University of Florida.

\section*{Data Availability}
The data underlying this article will be shared on reasonable request to the corresponding author.
\bibliographystyle{mnras}
\bibliography{references}








\bsp	
\label{lastpage}
\end{document}